\documentclass[%
 reprint,,bibtex,
superscriptaddress,
nofootinbib,
bibnotes,
 amsmath,amssymb,
 aps,
prl,
]{revtex4-1}
\usepackage{float}
\usepackage[dvipsnames]{xcolor}
\usepackage{pdfpages}
\usepackage{amsmath}
\usepackage{units}
\usepackage{graphicx}
\usepackage{dcolumn}
\usepackage{bm}
\usepackage{hyperref}
\usepackage{tabularx}
\usepackage[utf8]{inputenc}

\newcommand{\lk}{\left(}
\newcommand{\rk}{\right)}

\newcommand{\lK}{\left[}
\newcommand{\rK}{\right]}

\newcommand{\bk}{\mathbf k}
\newcommand{\bq}{\mathbf q}
\newcommand{\bb}{\mathbf b}
\newcommand{\bs}{\mathbf s}
\newcommand{\bu}{\mathbf u}
\newcommand{\br}{\mathbf r}
\newcommand{\bR}{\mathbf R}

\begin{document}

\preprint{APS/123-QED}

\title{Canonical Quantization of Crystal Dislocation and Electron-Dislocation Scattering in an
Isotropic Medium
}

\author{Mingda Li}
\email{mingda@mit.edu}
\affiliation{Department of Mechanical Engineering, Massachusetts Institute of Technology, Cambridge, MA 02139, USA}

\author{Wenping Cui}
\affiliation{Department of Physics, Boston College, Chestnut Hill, MA 02467, USA}

\author{M. S. Dresselhaus}
\affiliation{Department of Physics and Department of Electrical Engineering and Computer Science, Massachusetts Institute of Technology, Cambridge, MA 02139, USA}

\author{Gang Chen}
\email{gchen2@mit.edu}
\affiliation{Department of Mechanical Engineering, Massachusetts Institute of Technology, Cambridge, MA 02139, USA}

\date{\today}

\begin{abstract}
Crystal dislocations govern the plastic mechanical properties of materials but also affect the electrical and optical properties. However, a fundamental and quantitative quantum-mechanical theory of dislocation remains undiscovered for decades. Here we present an exactly solvable quantum field theory of dislocation, for both edge and screw dislocations in an isotropic medium by introducing a new quasiparticle “dislon”. With this approach, the electron-dislocation relaxation time is studied from electron self-energy which can be reduced to classical results. Moreover, a fundamentally new type of electron energy Friedel oscillation near dislocation core is predicted, which can occur even with single electron at present. For the first time, the effect of dislocations on materials’ non-mechanical properties can be studied at a full quantum field theoretical level. 
\begin{description}
\item[PACS numbers]
 03.70.+k, 61.72.Lk, 72.10.Fk.
\end{description}
\end{abstract}
\maketitle

Crystal dislocations are a basic type of one-dimensional topological defect in crystals \cite{1nabarro1967theory}. Since Volterra’s ingenious prototype in 1907 \cite{2volterra1907equilibre} and Taylor, Orowan and Polyani’s simultaneous formal introduction of such defects in 1934 \cite{3taylor1934mechanism,4Orowan,55polanyi1934lattice}, dislocations have been shown to have strong influences on materials, including the governing role in the plastic mechanical process, and widespread impact on materials’ thermal, electrical and optical properties etc \cite{1nabarro1967theory,6hirth1982theory}.
However, despite being one of the major driving forces in the development of modern metallurgy and material
sciences, a fundamental quantum-mechanical theory of dislocations is long overdue, partly due to the multivalued or discontinuous displacement field leading to a difficulty in going from a particular lattice model to an effective theory. The lack of a fundamental quantum-field theory of dislocations has impeded the study of the role of dislocations on non-mechanical material properties, limiting the usage of the terms “impurity” and “disordered systems” to point-defect related properties in many circumstances \cite{Ziman}.

In general, the theoretical attempts to study the role of a dislocation in material’s non-mechanical properties can be divided into at least 2 mainstream categories. One is classical scattering theory, where a dislocation is treated as a classical charged line \cite{7ng1998role,8weimann1998scattering,9jena2000dislocation,10look1999dislocation}. This allows one to study electron-dislocation scattering, but with obvious limitations when presented in a classical framework without taking into account of the displacement field a dislocation induces. The other is a geometrical approach, where a dislocated crystal is treated as a lattice in curved space as in general relativity \cite{11yavari2012riemann,12schmeltzer2012geometrical,13sansour2012generalized,14ruggiero2003einstein,Dislocation_GR}. This approach includes the displacement-field scattering but very problem-specific and mathematically cumbersome. Another promising approach is a gauge theory of a dislocation \cite{15kleinert1989gauge,16edelen2012gauge, Lazar}, which resembles quantized gauge theory in form but is limited only to classical elasticity field.

In this Letter, we provide an exact and mathematically manageable quantum field theory of a dislocation line, based on classical dislocation theory and a canonical quantization procedure. We find that in an isotropic medium, the exact Hamiltonian for both the edge and screw dislocations can be written as a new type of harmonic-oscillator-like Bosonic excitation along dislocation line, hence the name “dislon”. A dislon is similar to a phonon as lattice displacement with both kinetic and potential energy, but satisfying dislocation’s topological constraint $\oint_L d\bu= -\mathbf b$, where $\bb$ is the Burgers vector, $L$ is a closed contour enclosing the dislocation line, and $\mathbf u$ is the elastic displacement vector. For the first time, the scattering between an electron and the 3D displacement field induced by dislocation can easily be solved through many-body approach, with the topological constraint $\oint_L d\bu= -\mathbf b$ maintained all along.
 
\begin{figure}[h]
\centering
\includegraphics[width=0.85\columnwidth]{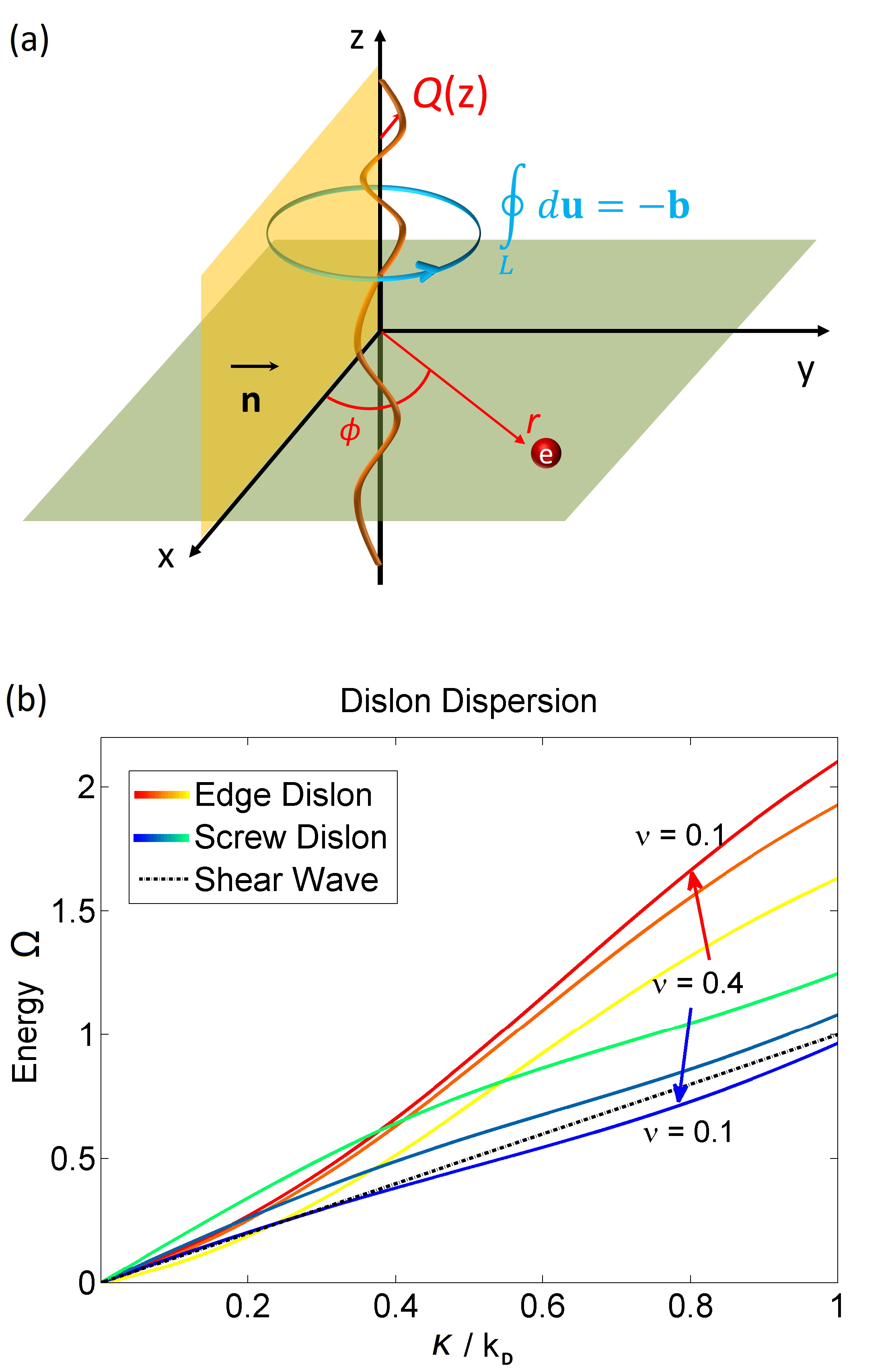}
\caption{(color online). (a) A long dislocation line along the $z$-direction vibrating within slip plane ($x$z) with $Q(z)$ the transverse displacement. Such vibration share similarities with phonons as it
is also quantized lattice vibration, but constrained by $\oint_L d\bu= -\mathbf b$
where $L$ is an arbitrary loop circling dislocation. An electron located at position $r$ will be scattered by dislocations. (b) Quantized vibrational excitation along dislocation line (“dislon”)
for both edge (hot-colors) and screw (cool-colors) dislocations at various Poisson ratios $\nu$. The classical shear wave is shown as a linear-dispersive black-dotted line, and the arrows indicate a decreasing trend of Poisson ratio for edge (red arrow) and screw (blue arrow) dislons, respectively.}
\label{Fig1}
\end{figure}

To begin with, defining $\bu(\bR)$ as the displacement at 3D position
vector $\bR$, its $i^{\rm th}$ component can be written
generically as \cite{17kosevich1986course, 18deWit1960249}
\begin{equation}\label{eq1}
u_i(\bR)=-c_{jklm}\int u_{lm}(\bR')\frac{\partial G_{ij}(\bR-\bR')}{\partial R_k}d^3\bR'
\end{equation}
where $i, j, k, l, m=1, 2, 3$ are Cartesian components, $c_{ijkl}=\lambda\delta_{ij}\delta_{kl}+\mu(\delta_{ik}\delta_{jl}+\delta_{il}\delta_{jk})$ is the elastic constant
tensor in an isotropic medium, $u_{lm}$ is the plastic distortion tensor, $G_{ij}$ is the Green’s function of the force equilibrium
equation whose Fourier transform can be written as \cite{18deWit1960249}
\begin{eqnarray}\label{eq2}
G_{ij}(\bk)&=&\frac{1}{\mu}\lK\frac{\delta_{ij}}{k^2}-\frac{\lambda+\mu}{\lambda+2\mu}\frac{k_ik_j}{k^4} \rK \nonumber\\
&=&\frac{1}{\mu}\lK\frac{\delta_{ij}}{k^2}-\frac{1}{2(1-\nu)}\frac{k_ik_j}{k^4} \rK
\end{eqnarray}
where $\lambda$ and $\mu$ are 1$^{\rm st}$ and 2$^{\rm nd}$ Lam\'e constants, respectively, and $\nu=\frac{\lambda}{2(\lambda+\mu)}$ is the Poisson ratio.

For a long, straight dislocation line extending along the $z$-direction and sitting at $(x_0,y_0)=(0,0)$, this dislocation line would vibrate within the slip plane ($xz$ plane) as an atomic motion, similar to phonons. Such vibration can be regarded as a precursor of the dislocation glide motion, and a direct generalization of long, straight static dislocation configuration. Defining $Q(z)$ to be the transverse displacement along the $x$ direction at position $z$, Eq. (\ref{eq1}) can be rewritten as \cite{18deWit1960249,19ninomiya1968dislocation} (See in Supplemental Material A)

\begin{equation}\label{eq3}
\resizebox{0.43 \textwidth}{!}{$
u_i(\bR)=\!u_i(\br,z)\!=\!-b_mc_{jklm}\int_{-\infty}^{+\infty}\! n_l \frac{\partial G_{ij}(\br,\!z\!-\!z')}{\partial R_k}Q(z')dz'$}
\end{equation}
where $\br=(x, y)$ and $\bR=(\br, z)$ are the 2D and 3D position vectors, $b_m$ is the $m^{\rm th}$ component of the Burgers vector and $\mathbf n$ is the direction perpendicular to the slip plane with the $l^{\rm th}$ component $n_l$. Mode-expanding the dislocation displacement as
\begin{equation}\label{eq4}
Q(z)=\sum_\kappa Q_{\kappa}e^{i\kappa z}
\end{equation}
where $\kappa$ is the wavenumber along the $z$-direction. Then the displacement $u(\bR )$ can be expressed as
\begin{equation}\label{eq5}
\resizebox{0.43 \textwidth}{!}{$
\!\!u_i(\bR)\!=\!\!\sum_\kappa f_i(\br;\kappa)e^{i\kappa z}Q_\kappa \!=\!\frac{1}{L^2}\sum_{\bs,\kappa}\!B_i(\mathbf{s};\kappa)e^{i\mathbf{s}\cdot\br+i\kappa z}Q_\kappa$}
\end{equation}
where $L$ is the system length, $\bs= ( k_x , k_y )$ is the 2D part of the 3D
wavevector $\bk = (\bs ,\kappa )$ in Eq. (\ref{eq2}), and
$B_i(\mathbf{s};\kappa)=\int f_i(\br;\kappa) e^{-i\bs\cdot\br}d^2\br$ is the 2D Fourier transform of
$f_i(\br;\kappa)$ which can be obtained as 
\begin{equation}\label{eq6}
B_i(\mathbf{s};\kappa)=+i\mu b_m n_l\lK k_mG_{il}(\bk)+k_lG_{im}(\bk) \rK
\end{equation}
by comparing Eq. (\ref{eq3}) to Eq. (\ref{eq5}). The classical 3D kinetic
and potential energy of such a vibrating dislocation can be
written as a 1D effective Hamiltonian \cite{19ninomiya1968dislocation}
\begin{equation}\label{eq7}
\resizebox{0.43 \textwidth}{!}{$
H=T+U=\frac{L}{2}\sum_\kappa m(\kappa)\dot{Q}_\kappa\dot{Q}^*_\kappa+\frac{L}{2}\sum_\kappa \kappa^2 K(\kappa)Q_\kappa Q^*_\kappa$}
\end{equation}
where $m(\kappa)$ and $\kappa^2K(\kappa)$ are identified as the classical linear mass density and tension, respectively, and can be written as \cite{19ninomiya1968dislocation} (See also in Supplemental Material B)
\begin{equation*}
\resizebox{0.48 \textwidth}{!}{$m(\kappa)\!=\!\frac{\rho}{4\pi^2}\int \! d^2\mathbf{s}\frac{1}{k^4}\lK\lk\mathbf{b}\cdot \bk\rk^2\!+\!b^2\lk\mathbf{n}\cdot \bk \rk^2
\!+\!\frac{4\nu-3}{\lk 1-\nu\rk^2}\frac{\lk \mathbf{n}\cdot \bk\rk^2\lk\mathbf{b}\cdot \bk \rk^2}{k^4} \rK$}
\end{equation*}
\begin{equation*}
\resizebox{0.48 \textwidth}{!}{$\kappa^2K(\kappa)\!=\!U_0\!-\!\frac{\mu}{4\pi^2}\int \! d^2\mathbf{s}
\lK\frac{\lk\mathbf{b}\cdot \bk\rk^2\!+\!b^2\lk\mathbf{n}\cdot \bk \rk^2}{k^2}\!-\!\frac{2}{1-\nu}\frac{\lk \mathbf{n}\cdot \bk\rk^2\lk\mathbf{b}\cdot \bk \rk^2}{k^4} \rK$}
\end{equation*}
where $U_{0}$ is a $\kappa$-independent zero-point-energy term depending on the dislocation type. 

Now imposing a canonical quantization condition,
\begin{equation}\label{eq8}
Q_\kappa =Z_\kappa \lK a_\kappa+a^+_{-\kappa} \rK,\quad P_\kappa =\frac{i\hbar}{2Z_\kappa} \lK a^+_\kappa-a_{-\kappa} \rK
\end{equation}
where $P_\kappa=\frac{\partial\mathcal{L}}{\partial \dot{Q_\kappa}}=Lm(\kappa)\dot{Q}^*_\kappa$, $Z_\kappa=\sqrt{\frac{\hbar}{Lm(\kappa)\omega(\kappa)}}$, then the classical dislocation Hamiltonian Eq. (\ref{eq7}) is quantized as (See Supplemental Material C)
\begin{equation}\label{eq9}
H_D=\sum_{\kappa}\hbar \omega(\kappa)\lK a^+_\kappa a_\kappa +\frac{1}{2}\rK
\end{equation}
with eigenfrequency $\omega(\kappa)=\kappa\sqrt{K(\kappa)/m(\kappa)}$ , which has the form as a collection of non-interacting Bosonic excitations. Despite the observation that such an excitation shares the similarity with the phonon excitation as a type of quantized lattice vibration, the topological constraint here $\oint_L d\bu= -\mathbf b$ leads to a different excitation quantum along the dislocation line, which may suitably be called a “dislon”, to distinguish the dislon from a non-interacting phonon. In particular,  by imposing the in-plane Debye cutoff $k_D$, the dislon dispersion relation for an edge dislocation $(\bb\bot z, b=b_x )$ and a screw dislocation $(\bb\| z, b=b_z )$, can separately be written in a closed-form as
\refstepcounter{equation}\label{eq10}
\begin{equation}
\resizebox{0.41 \textwidth}{!}
{$w_E(\kappa)= v_s\kappa\!\sqrt{\frac{C_1\mathrm{log}\lk \frac{k_D^2}{\kappa^2}+1 \rk +1-C_2\frac{\kappa^2}{k_D^2+\kappa^2}}{(1+C_3)\mathrm{log}\lk 1+\frac{k_D^2}{\kappa^2}\rk-\frac{k_D^2}{k_D^2+\kappa^2}-C_3\frac{k_D^2(3k_D^2+2\kappa^2)}{2(k_D^2+\kappa^2)}}}\tag{10a}\label{eq10a}$}
\end{equation}

\begin{equation}\label{eq10b}
\resizebox{0.41 \textwidth}{!}{
$w_S(\kappa)= v_s\kappa\!\sqrt{\frac{C_4\mathrm{log}\lk \frac{k_D^2}{\kappa^2}+1 \rk -C_4+4C_2\frac{\kappa^2}{k_D^2+\kappa^2}}{\frac{k_D^2}{2(\kappa^2+k_D^2)}+\frac{1}{2}\mathrm{log}\lk 1+\frac{k_D^2}{\kappa^2} \rk+2C_3\frac{k_D^4}{\lk k_D^2+\kappa^2\rk^2}}}\tag{10b}$}
\end{equation}
where $v_s=\sqrt{\mu/\rho}$ is the shear velocity, and the four coefficients are
$C_1=\frac{1-2\nu}{2(1-\nu)}$, $C_2=\frac{1}{4(1-\nu)}$, $C_3=\frac{4\nu-3}{8(1-\nu)^2}$ and $C_4=\frac{1+\nu}{2(1-\nu)}$. The dislon dispersion at various $\nu$ values and keeping constant $v_s = 1$ are plotted in Fig. \ref{Fig1}(b), where the classical shear wave, or equivalently the transverse acoustic phonon mode $\omega(\kappa)=v_s\kappa$ (black-dotted line) serves as a pre-factor in the quantum-mechanical version of the dislocation excitation in Eq. (\ref{eq10}). The higher excitation energy of the edge dislon than the screw dislon is reasonable, since even in a pure classical picture, the edge dislocation energy density is higher than the screw dislocation by a factor of $1/(1-\nu)$ \cite{20hull2001introduction}; due to the different zero-point-energy, they have opposite trends at various $\nu$ values (red and blue arrows in Fig. \ref{Fig1}b).
\begin{figure}
\centering
\includegraphics[width=1.00\columnwidth]{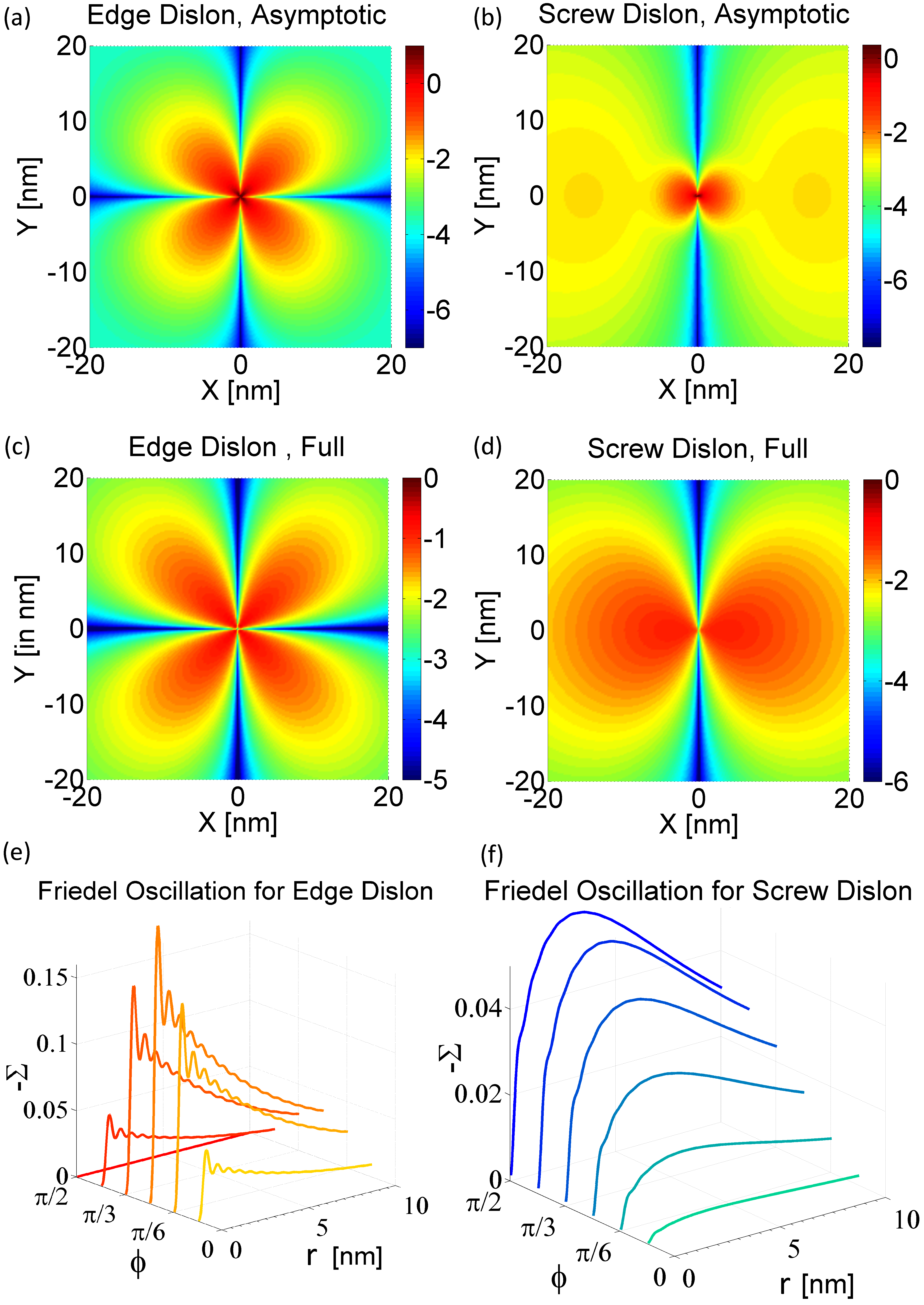}
\caption{(color online). The self-energy of an electron away from a dislocation line with a dislocation core $(x_0,y_0)=(0,0)$. The ratio between the self-energy $|\Sigma|$ and electron energy $\varepsilon_p$ on a logarithmic scale, as a function of the 2D coordinate $\br =( x,y)$ for edge (a, c) and screw (b, d) dislocations. The self-energy decays fast away from the dislocation core. Compared with the asymptotic exponential decay behavior (a, b), the full coupling constants (c, d) indeed reveal an exotic Friedel oscillation, which is anisotropic and can occur with only single electron at present. This can be seen more clearly on linear scale waterfall plots (e, f). The much more drastic oscillation for edge dislocation (e) than screw dislocation (f) is caused by dilatation effect and resulting electrostatic potential.}
\label{Fig2}
\end{figure}

To study the electron-dislocation interaction from a full $2^{\rm nd}$-quantization level, we start from a generic electron-ion interaction Hamiltonian \cite{21bruus2004many,22mahan2013many}
\begin{equation}\label{eq11}
H_{e-ion}=\int d^3\bR \rho_e(\bR)\sum_{j=1}^N\nabla_\bR V_{ei}(\bR-\bR^0_j)\cdot \bu_j\tag{11}
\end{equation}
where $\rho_e$ is the electron charge density, $\bR^0_j$ is the equilibrium position of the ion with label $j$, and sum on j is over all $N$ ions in the crystal. By expanding the electron-ion interaction potential $V_{ei}$ as 
\begin{equation}\label{eqex}
V_{ei}(\bR-\bR^0_j)=\frac{1}{V}\sum_{q\in 1\mathrm{BZ} }\sum_{\mathbf{G}}V_{\bq+\mathbf{G}}e^{i(\mathbf{G}+\bq)\cdot(\bR-\bR^0_j)}\tag{12}
\end{equation}
and using Eqs. (\ref{eq5})-(\ref{eq8}), and moreover by assuming a non-Umklapp normal process $(\mathbf{G}=0)$,
the electron-dislocation interaction Hamiltonian Eq. (\ref{eq11})
can be re-written as (see Supplemental Material D)
\begin{equation}
\begin{aligned}\label{eq12}
H_{el-dis}=&-\frac{N}{V}\frac{1-2\nu}{1-\nu}\sum_\bq \rho(\bq)eV_\bq\nonumber\frac{(\bb\cdot\bq)(\mathbf{n}\cdot\bq)}{L^2q^2}\times\\
&\sqrt{\frac{\hbar}{2Lm(\kappa)\omega(\kappa)}}(a_\kappa+a^+_{-\kappa})
\end{aligned}\tag{13}
\end{equation}
where the screened Coulomb potential $V_\bq=\frac{4\pi Z e}{q^2+k^2_{TF}}$ with $k_{\rm TF}$
defined as the Thomas-Fermi screening wavenumber, $\rho(\bq)$ is the Fourier-transformed electron number density. At $\nu = 1/2$ both Eqs. (\ref{eq9}) \& (\ref{eq12}) vanish due to the important pre-factor $1-2\nu$. This is reasonable since at $\nu = 1/2$ the system demonstrates only elasticity without plasticity (such as rubber), resulting in both vanishing dislon excitation and quantized electron-dislocation interaction. For a single electron located at $\bR=(\br,z)=(r\cos\phi,r\sin\phi,z)$,
we have $\rho(\bq)=e^{i\mathbf{s}\cdot\br+i\kappa z}$, and Eq. (\ref{eq12}) can be further simplified as
\begin{equation}\label{eq13}
H_{e-dis}=\frac{1}{\sqrt{L}}\sum_\kappa e^{i\kappa z}{M}_{\bb,\br}(\kappa)(a_\kappa+a_{-\kappa}^+)\tag{14}
\end{equation}
with the single-electron position-dependent coupling constant ${M}_{\bb,\br}(\kappa)$ written as
\begin{equation}
\begin{aligned}\label{eq14}
&{M}_{\bb,\br}(\kappa)=\frac{Ne^2}{V}\frac{1-2\nu}{1-\nu}\sqrt{\frac{\hbar}{2m(\kappa)\omega(\kappa)}}\nonumber\\
&\times\int_0^{k_D}ds s^2\frac{b_s  s J_2(rs)sin(2\phi)-2ib_z\kappa J_1(rs)\sin\phi}{(s^2+\kappa^2+k_{TF}^2)(s^2+\kappa^2)}
\end{aligned}\tag{15}
\end{equation}
where $J_n(rs)$ is the $n^{\rm th}$ order Bessel function of the $1^{\rm st}$ kind. One promising feature here is that the 3D electron-displacement interaction in Eq. (\ref{eq11}) can be rewritten in an effective 1D electron-dislon interaction as Eq. (\ref{eq13}), greatly simplifying the calculation of dislocation's influence to electron motion. In particular, in the non-screening, stiff-solid limit $k_{\rm TF}\rightarrow 0$ and $k_D \rightarrow \infty$ , and in $r \rightarrow \infty$ limit where an electron is far apart from a dislocation core, the asymptotic coupling constants have a closed-form:
\begin{align}
&|M_{\substack{edge\\r \rightarrow \infty}}(\kappa)|^2\tag{16a}\label{eq15a}
\\=&\lk\frac{Ne^2}{8V}\rk^2\lk\frac{1-2\nu}{1-\nu}\rk^2\frac{\hbar \pi b^2(2\kappa r-3)^2}{m(\kappa)\omega(\kappa)\kappa r}\sin^2(2\phi)e^{-2\kappa r}\nonumber\\\nonumber\\
&|M_{\substack{screw\\r \rightarrow \infty}}(\kappa)|^2\tag{16b}\label{eq15b}
\\=&\lk\frac{Ne^2}{8V}\rk^2\lk\frac{1-2\nu}{1-\nu}\rk^2\frac{4\hbar \pi b^2(2\kappa r-1)^2}{m(\kappa)\omega(\kappa)\kappa r}\sin^2(\phi)e^{-2\kappa r}\nonumber
\end{align}
which shows an exponential decay of the electron-dislocation coupling strength at long distance. 

In addition, we could write down the Feynman rule for electron-dislocation scattering accordingly from the coupling constants (See Supplemental Material E):
Internal dislon line
$|{M}_{\bb,\br}(\kappa)|^2D^{(0)}(\kappa,i\omega_m)$,
with $D^{(0)}(\kappa,i\omega_m)=-\frac{2\omega_\kappa}{\omega_m^2+\omega_\kappa^2}$ 
and $\omega_m=2m\pi/\beta$($\beta=1/(k_{B}T)$), and other Feynman rules unchanged. Therefore, to one-loop correction, in which case an electron emits and re-absorbs a virtual dislon, we could compute the position-dependent electron self-energy as follows:
\begin{align}\label{eq16}
&\Sigma^{(1)}(\br,\mathbf{p},E)=\int \frac{d\kappa}{2\pi}|{M}_{\bb,\br}(\kappa)|^2\tag{17}\\
&\times\lK
\frac{n_B(\omega_\kappa)+n_F(\varepsilon_{\mathbf{p}+\kappa})}{E-\varepsilon_{\mathbf{p}+\kappa}+\omega_\kappa+i\delta}\!+
\frac{n_B(\omega_\kappa)+1-n_F(\varepsilon_{\mathbf{p}+\kappa})}{E-\varepsilon_{\mathbf{p}+\kappa}-\omega_\kappa+i\delta}
\rK\nonumber
\end{align}
We take germanium as a prototype example since it has isotropic bands, but we do not intend to compute real material. At zero temperature and using reasonable values of elastic parameters of germanium \cite{23claeys2008extended}( $b =\unit[0.4]{nm} $, $\rho= \unit[5.3]{g/cm^3}$, $\mu = \unit[67]{GPa}$, $\nu = 0.28$, cutoff $\kappa_{min} =\unit[0.05]{nm^{-1}}$ and $\kappa_{max} =\unit[10]{nm^{-1}}$), the electron on-shell self-energy is plotted in Fig. \ref{Fig2}, for edge (Fig. \ref{Fig2} a, c, e) and screw (Fig. \ref{Fig2} b, d, f) dislocations, with
full coupling constants Eq. (\ref{eq14}) (Fig. \ref{Fig2} a, b) and asymptotic form Eqs. (\ref{eq15a}, b) (Fig. \ref{Fig2} c, d). Note $\Sigma< 0$ in all cases, indicating a reduction of the electron energy, hence an increase of the electron effective mass when being scattered by a dislocation, similar to electron-phonon scattering. The 4-fold self-energy symmetry for the edge dislocation and 2-fold symmetry for the screw dislocation is also reasonable- the classical displacement field distributions $\bu(\bR )$ have 2-fold and 1-fold symmetry from sinusoidal terms, respectively, while the energy $\sim |\bu(\bR )|^2$ doubles the symmetry. The most prominent feature is that the electron self-energy shows an anisotropic Friedel oscillation behavior (Fig. \ref{Fig2} e, f). Since a single dislocation could be treated as a classical 1D charged line, such a quantum-mechanical oscillation may not appear surprising as a direct generalization to a 0D point-defect Friedel oscillation. However, what is striking is that such an oscillation can occur with the presence of only one electron, since it is the oscillation of the single-electron self-energy instead of charge density oscillation which gives rise to the traditional Friedel oscillation. Another feature is that the oscillation caused by an edge dislocation (Fig. \ref{Fig2} e) is much more drastic than a screw dislocation (Fig. \ref{Fig2} f). This can be understood from the distinct electrostatic effect contributing to Friedel oscillation. For an edge dislocation there is finite inhomogeneous dilatation $\Delta=-\frac{b}{2\pi}\frac{1-2\nu}{1-\nu}\frac{sin\theta}{r}$, leading to a compensating electrostatic potential to reach uniformly distributed Fermi energy and equilibrium, while for a screw dislocation, linear elasticity gives no dilatation hence no electrostatic effect emerges \cite{1nabarro1967theory}. The observation of the predicted self-energy's single-electron Friedel oscillation may provide strong evidence of the existence of dislon and the quantum nature of dislocations. 

\begin{figure}
\centering
\includegraphics[width=1.00\columnwidth]{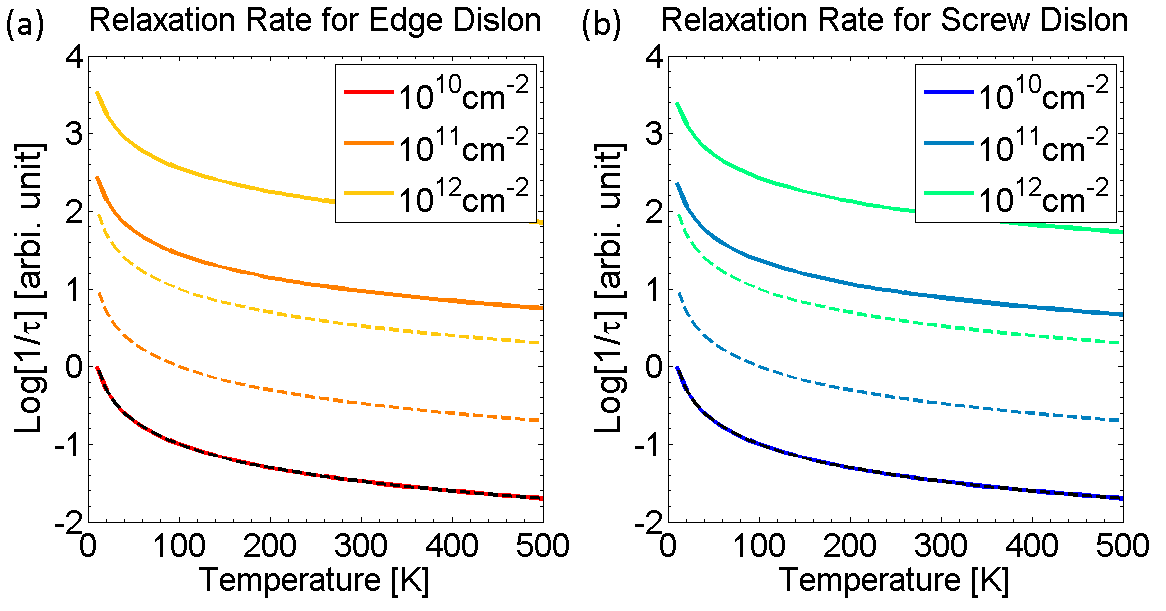}
\caption{(color online). Comparison of the normalized logarithmic relaxation rate
using the classical result \cite{24dexter1952effects}(dashed lines) and Eq. (\ref{eq16})(solid lines).}
\label{Fig3}
\end{figure}
To test the power of this theoretical framework, we compare the electron-dislocation scattering relaxation rate $\frac{1}{\tau_{dis}(\br)} \propto \mathrm{Im}\Sigma(\br,\mathbf{p},\varepsilon_p)$ from Eq. (\ref{eq16}) to the classical results with empirical deformation-potential parameters \cite{24dexter1952effects}. Despite different methods, our one-loop result Eq. (\ref{eq16}) shares an identical pre-factor $\lk\frac{1-2\nu}{1-\nu} \rk^2$ and the same temperature dependence with the classical relation $\frac{1}{\tau_{dis}(\br)} \propto \frac{N_{dis}}{T}\lk\frac{1-2\nu}{1-\nu} \rk^2$, but has a stronger capability to compute the position and energy dependent relaxation time. Assuming the average dislocation-electron distance $\bar{r} = \sqrt{1/ N_{dis}}$ , the comparison (normalized at $\unit[10^{10}]{cm^{-2}}$) of the relaxation rate is plotted in Fig. \ref{Fig3} and shows very similar trends.

To summarize, we have demonstrated a quantized Bosonic excitation along a crystal dislocation, “dislon”,
whose excitation spectra are obtained in closed-form in an isotropic medium. Such a framework allows the study of classical electron static-dislocation scattering at a full dynamical many-body quantum-mechanical level, and is expected to greatly facilitate the study of the effects of isolated dislocations on the electrical properties of materials because of the case with which the effects of an isolated dislon can be incorporated into existing theories without loss of rigor. In fact, with a fully quantized dislocation, it can be shown a decades-long debate of the nature of dislocation-phonon interaction-whether static strain scattering or dynamic fluttering dislocation scattering- shares the same origin as phonon renormalization \cite{quasi_phonon}. What's more, since the dislon is a type of Bosonic excitation, the dislon may also couple 2 electrons to form Cooper pair, becoming an extra contributor to superconductivity besides a phonon. This may seem counterintuitive since dislocations as defects would only shorten the electron mean free path and lead to weakening of superconducting phenomena, yet early experimental evidence did show a sample annealing temperature-dependent superconducting transition temperature $T_c$ \cite{25Hauser1961Annealing}, whereby different samples have identical stoichiometry but different dislocation densities, and a slight increase of $T_c$ under plastic deformation in another experiment \cite{26Tc}, indicating a more profound role that dislocations may play in superconductivity, but further studies of this phenomenon is now under more detailed investigation. 

\acknowledgements
M.L. would thank helpful discussions with Prof. Hong Liu, and suggestions by Jiawei Zhou, Shengxi Huang, and Maria Luckyanova. M.L., M.S.D. and G.C. would like to thank support by S$^{3}$TEC, an EFRC funded by DOE BES under Award No. DE-SC0001299/DE-FG02-09ER46577.

\bibliography{dislon}
\end{document}